\documentclass[fleqn,12pt]{article}
\usepackage{espcrc1,epsfig}

\usepackage{graphicx}
\usepackage[figuresright]{rotating}

\def\Pom{{\bf I\!P}}

\def\lsim{\mathrel{\rlap{\lower4pt\hbox{\hskip1pt$\sim$}}
    \raise1pt\hbox{$<$}}}         

\def\gsim{\mathrel{\rlap{\lower4pt\hbox{\hskip1pt$\sim$}}
    \raise1pt\hbox{$>$}}}         

\newcommand{\be}{\begin{equation}}
\newcommand{\ee}{\end{equation}}
\newcommand{\bea}{\begin{eqnarray}}
 
\newcommand{\eea}{\end{eqnarray}}

\newcommand{\ba}{\begin{array}}
\newcommand{\ea}{\end{array}}

\newcommand{\bDelta}{\mbox{\boldmath $\Delta$}}

\newcommand{\br}{{\bf r}}

\newcommand{\AmS}{{\protect\the\textfont2
  A\kern-.1667em\lower.5ex\hbox{M}\kern-.125emS}}

\title{Unifying aspects of polarization of
vector mesons from hard production in DIS and at Tevatron}

\author{N.N. Nikolaev \address[IKPFZJ]{Institut f. Kernphysik, Forschungszentrum
J\"ulich\\ D-52425 J\"ulich, Germany\\
and\\
L.D. Landau Institute for Theoretical Physics\\ 142432 Chernogolovka, Russia}%
        \thanks{Supported partly by the grant INTAS-597/34094 and DAAD-NSC(Taiwan) 
joint research grant S014160}}
\begin{document}

\maketitle

\begin{abstract}
Virtual photons have both the transverse and longitudinal polarization. In 
the inclusive DIS the impact of longitudinal photons,
quantified by $R=\sigma_L/\sigma_T \sim 0.2$, is marginal, but
in diffractive DIS with excitation of small mass hadronic states 
or exclusive diffraction into vector mesons, QCD predicts
$R \gg 1$ in agreement with the experiment. After a brief review
of the modern status of QCD theory of diffractive vector meson
production I argue that longitudinally polarized gluons give
rise to a large longitudinal polarization of the prompt $J/\Psi$
and $\Psi'$ observed at the Tevatron.
\end{abstract}

\section{Introduction}

In DIS incident leptons serve as 
a source of virtual photons and experimentally one studies a
virtual photoproduction of various hadronic states. By virtue
of the optical theorem, 
the inclusive DIS structure functions are related to the imaginary part of 
an amplitude of diagonal, $Q^2_f = Q_{in}^2 =Q^2$, forward virtual
Compton scattering (CS)
$
\gamma^{*}_{\mu}(Q^2_{in})p\to \gamma^{*}_{\nu}(Q_f^2)p'$,
which for the reason of vanishing $(\gamma^*,\gamma^*)$
momentum transfer happens to be diagonal in the photon 
helicities, ${\nu}={\mu}$. While real photons are 
transverse ones, i.e., have only circular 
polarizations, $\mu = \pm 1$,  
virtual photons radiated by leptons have also the longitudinal 
polarization, which in the 
scaling limit equals  
\be
\epsilon_L = {2(1-y) \over 2(1-y)+y^2},
\label{eq:1.3}
\ee
where $y$ is a fraction of the beam lepton energy taken away by the photon,
so that the photoabsorption cross section measured in 
the inclusive DIS equals
$\sigma = \sigma_T+\epsilon_L \sigma_L$. The effect  of 
longitudinal photons,
quantified by $R_{DIS}=\sigma_L/\sigma_T \sim 0.2$, is marginal, though.

Keep the virtuality of the initial photon, $Q^2=Q_{in}^2$, fixed.
By analytic continuation to $Q_{f}^{2}=0$ one obtains DVCS, the 
still further continuation to $Q_{f}^{2}=-m_{V}^2$ transforms CS 
into the diffractive vector meson (VM)
production
$ 
\gamma^{*}_{\mu}(Q^2)p\to V_{\nu}(\bDelta)p'(-\bDelta)\, ,
$ 
which is accessible experimentally also at finite  $(\gamma^* V)$ momentum
transfer $\bDelta$. Vector mesons have the 
three polarization states, $\nu = \pm 1,0$. The decays of VM's are
self-analyzing which allows to reconstruct
the full set of helicity amplitudes $A_{\nu\mu}$ and probe the 
production mechanism in full complexity. The crucial point
about diffractive excitation of VM and small mass continuum
is that they are entirely dominated by $\sigma_{L}$ \cite{NNZscan,GNZsigmaL}.

In this report I review first the spin phenomena in diffractive 
exclusive vector meson production. The new numerical
results reported here were obtained in collaboration with Igor Ivanov
\cite{Igor}. Then I turn to the r\^ole of longitudinally polarized 
gluons in inclusive production of vector mesons in hadronic production
and shall argue that they are a natural source of large longitudinal
polarization of the $J/\Psi$ and $\Psi'$ mesons as discovered ed at Tevatron
\cite{CDF}.
This observation is from an ongoing collaboration work with Pauchy Hwang, Igor
Ivanov and Wolfgang Sch\"afer \cite{JPsiPsiPrim}.

\section{Color dipole factorization,  $(Q^{2}+m_{V}^2)$ scaling
and spin dependence of vector meson production}

CS and diffractive VM production at small-$x$ are 
best described in color dipole (CD) factorization \cite{NNZscan}, in
which  
$
A_{\nu\mu}=\Psi^{*}_{\nu,\lambda\bar{\lambda}}\otimes A_{q\bar{q}}\otimes
\Psi_{\mu,\lambda\bar{\lambda}}
$
where $\lambda,\bar{\lambda}$ stands for $q,\bar{q}$ helicities,
$\Psi_{\mu,\lambda\bar{\lambda}}$ is the wave function (WF) of the $q\bar{q}$ 
Fock state of the photon or VM. The  $q\bar{q}$-proton 
scattering kernel $A_{q\bar{q}}$ is proportional to color dipole cross section,
does not depend on, and conserves exactly, the $q,\bar{q}$ helicities. 
For small dipoles, the CD cross section described by the  two-gluon 
QCD pomeron exchange is manifestly related to the gluon SF 
of the target ($A\approx 10$ follows from properties of of Bessel functions 
\cite{NZglue}), 
\be
\sigma(x,\br)\approx {\pi^2 \over 3}r^2 \alpha_{S}({A\over r^2})
G(x,{A\over r^2})\, . 
\label{eq:2.1}
\ee

\begin{figure}[!htb]
   \centering
   \epsfig{file=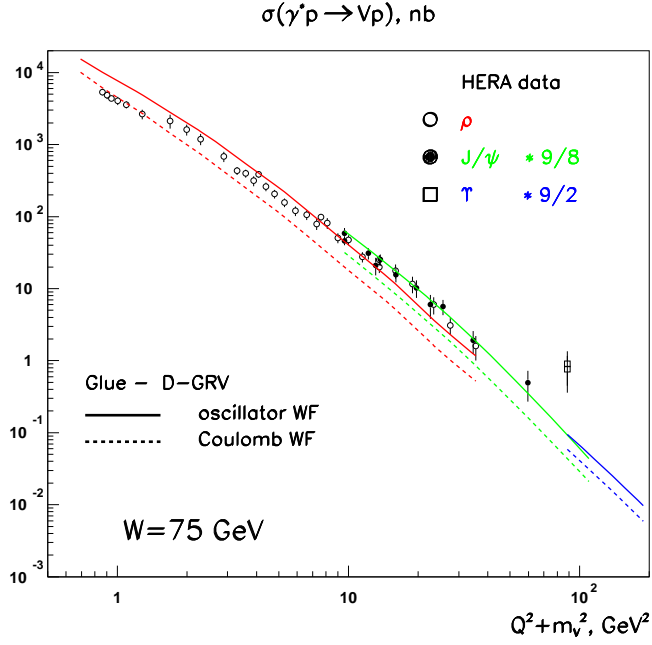,width=10cm}
\vspace{-0.5cm}
\caption{The test of the $(Q^2+m_V^2)$ scaling. The divergence of the 
solid and dashed curves indicates the sensitivity to the WF of the VM.
The experimental data are from HERA \cite{HERAdata}.} 
\label{Figure1}
\end{figure}

In exclusive production of the VM one swaps in the final
state the pointlike photon, whose $q\bar{q}$ WF  is
singular at $r\to 0$ \cite{NZ91}, for the finite-size  
VM with WF which is smooth at $r\to 0$. As a result, while  DIS probes 
$\sigma(x,\br)$ CD in a broad
range of ${1 \over AQ^2} \lsim r^2 \lsim 1 $ fm$^2$ \cite{NZHERA},
the diffractive VM 
production probes $\sigma(x,\br)$, and the VM WF, at a 
scanning radius  \cite{NN92,NNZscan} 
\be
r\sim r_{S}= {6 \over \sqrt{Q^2 + m_{V}^2}}\, .
\label{eq:2.2}
\ee
Regarding the spin dependence of diffractive VM, the fundamental
point 
is that the sum of quark and antiquark helicities equals helicity of 
neither the photon nor vector meson. If for the 
nonrelativistic massive quarks, $m_{f}^{2} \gg Q^{2}$ the only allowed 
transition is $\gamma^{*}_{\mu} \to q_{\lambda} +\bar{q}_{\bar{\lambda}}$ 
with $\lambda +\bar{\lambda}=\mu$. In the relativistic case transitions
of transverse photons $\gamma^{*}_{\pm}$ into the $q\bar{q}$ state with 
$\lambda +\bar{\lambda}=0$, 
in which the helicity of the photon is transferred to the $q\bar{q}$ orbital 
 momentum, are equally allowed. Consequently, in QCD the 
$s$-channel helicity non-conserving (SCHNC) transitions
$\gamma^{*}_{\pm} \to (q\bar{q})_{\lambda +\bar{\lambda}=0} \to
\gamma^{*}_{L} ~~~{\rm and}~~~
\gamma^{*}_{\pm} \to (q\bar{q})_{\lambda +\bar{\lambda}=0 }\to
\gamma^{*}_{\mp} $ 
are allowed \cite{NPZLT,KNZ98} and SCHNC persists at small $x$
despite the exact conservation of the helicity of quarks in  
$q\bar{q}$-target scattering. This argument 
for SCHNC does not require the applicability of pQCD.
Furthermore, the leading  contribution to the proton structure function
comes entirely from SCHNC transitions of transverse photons - the
fact never mentioned in textbooks.

Still another fundamental point is that the vertex of the SCHC 
transition
$\gamma^*_L \to (q\bar{q})_{\lambda +\bar{\lambda}=0}$ is proptional
to $Q$, which entails \cite{NNZscan}
\be
R= {\sigma_L(\gamma^*_L p\to V_L p) \over 
\sigma_T(\gamma^*_T p\to V_T p)} \sim { Q^2 \over m_V^2} >> 1
\label{eq:4.2}
\ee
for diffractive VM's. As was first noticed 
in \cite{NNZscan}, a numerical analysis with realistic 
soft WF gives values of $R$ substantially smaller than a crude
estimate (\ref{eq:4.2}).

The three fundamental consequences of (\ref{eq:2.1}), (\ref{eq:2.2})
and (\ref{eq:4.2}) are:
\begin{itemize}
\item
the VM production probes \cite{NNZscan} 
the gluon SF of the target at the hard scale
$\overline{Q}^2 \approx$ (0.1-0.25)$* (Q^2 + m_{V}^2)$ 
and $x=0.5(Q^2+m_{V}^2)/(Q^2+W^2)$,
\item
 after factoring out the charge-isospin factors all 
VM production cross section follow a universal function of $\overline{Q}^2$,
i.e. there is $(Q^2 + m_{V}^2)$ scaling \cite{NNZscan}, see fig.~1, the
same scaling holds also
for the effective intercept $\alpha_{\Pom}(0)-1$ of the energy
dependence of the production amplitude, see fig.~2,
\item
the contribution to the diffraction slope $B$ from the $\gamma^* \to V$
transition vertex decreases $\propto r_{S}^2$ exhibiting again the 
$(Q^2 + m_{V}^2)$ scaling \cite{NZZslope}, see fig.~2.

\end{itemize}
\begin{figure}[!htb]
\vspace{-0.5cm}   
\centering
   \epsfig{file=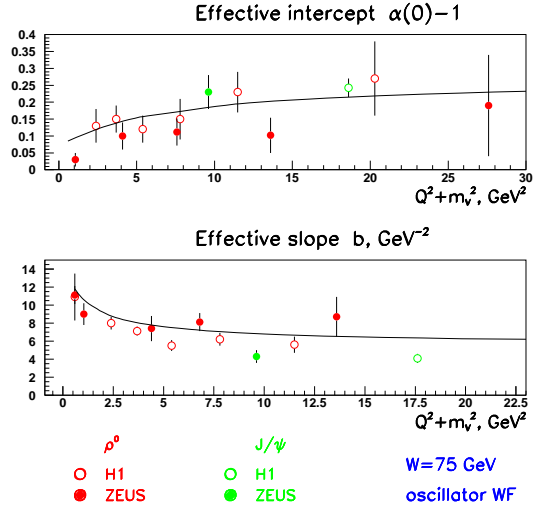,width=8cm}
\vspace{-0.5cm}
\caption{The $(Q^2+m_V^2)$ scaling of the effective intercept and diffraction slope
\protect{\cite{Igor}}}
\label{Figure2}
\end{figure}

The agreement between theory and experiment \cite{HERAdata}
 is good, although there remains certain sensitivity
to not so well known WF of VM's which
can not be eliminated at the moment, see also below.
The theoretical calculations are based on the differential 
glue in the proton found in \cite{INdiffglue}

We emphasize that SCHNC helicity flip only is possible due to 
the transverse and/or longitudinal Fermi motion of quarks and
is extremely sensitive to spin-orbit coupling in the
vector meson, I refer for details to \cite{KNZ98,IN99}.
The consistent analysis of production of 
$S$-wave and $D$-wave vector mesons is presented only
in \cite{IN99}.  The dominant 
SCHNC effect in vector meson production is the interference 
of SCHC $\gamma^{*}_{L} \to V_L$ and SCHNC $\gamma^{*}_{T}\to V_L$ production,
i.e., the element $r_{00}^{5}$ of the vector meson spin
density matrix. The overall agreement
between our theoretical estimates 
\cite{Igor} of the 
spin density matrix $r_{ik}^{n}$ for
diffractive  $\rho^{0}$ assuming pure $S$-wave in the $\rho^{0}$-meson
and the ZEUS \cite{ZEUSflip} and H1 \cite{H1flip} experimental data
is very good, there is a clear evidence for  $r_{00}^{5}\neq 0$, see
fig.~3.

\begin{figure}[!htb]
   \centering
   \epsfig{file=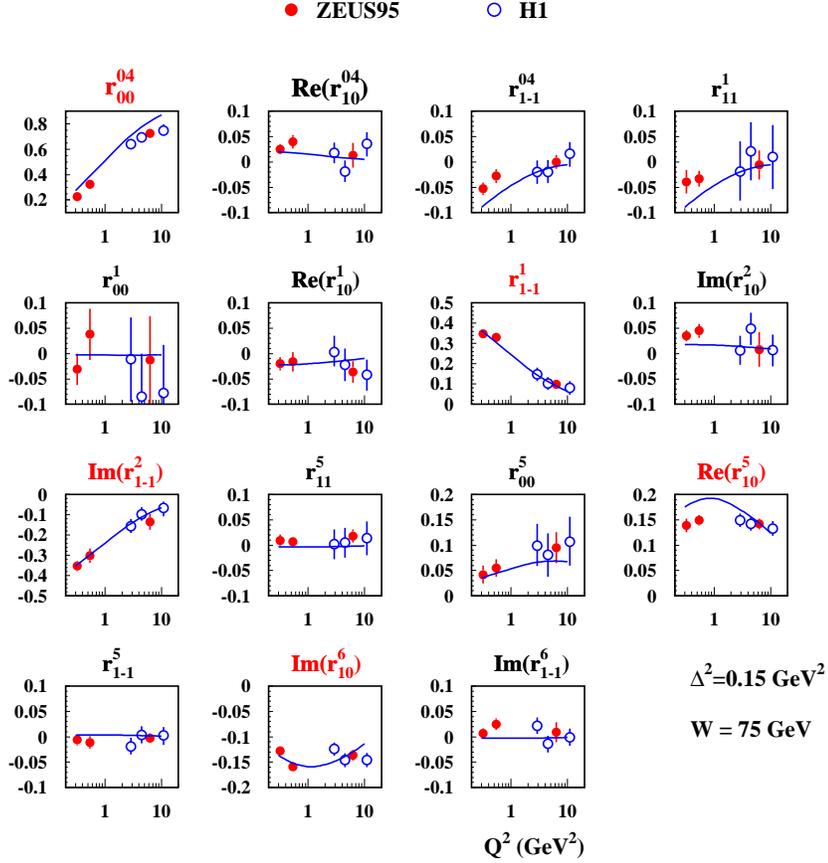,width=12cm}
\vspace{-0.5cm}
\caption{Predictions for the spin density matrix in the $\rho^{0}$
production vs. the experimental data from HERA \protect{\cite{ZEUSflip,H1flip}}.}
\label{Figure3}
\end{figure}

\begin{figure}[!htb]
\vspace{-.3cm}
   \centering
   \epsfig{file=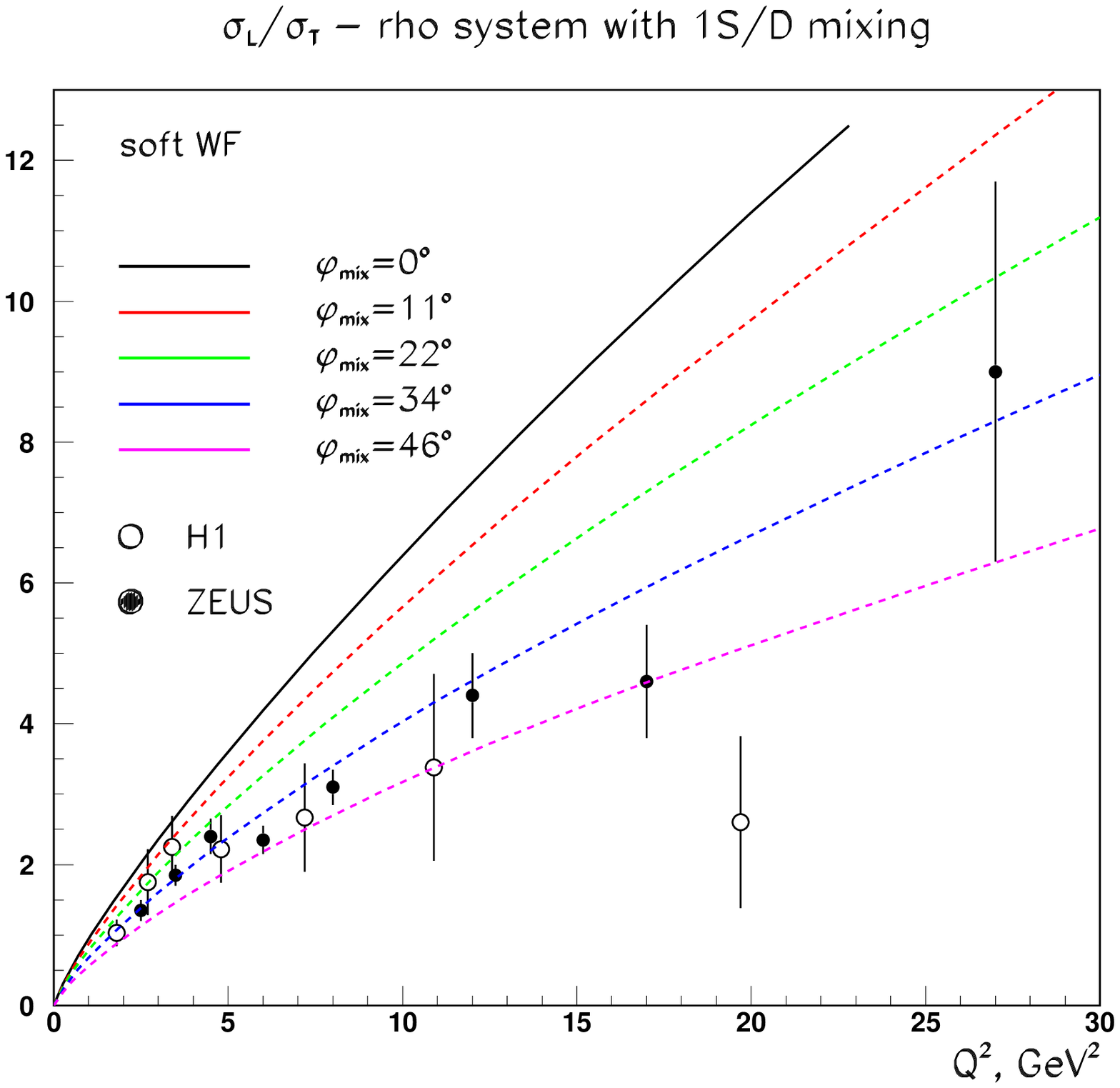,width=9cm}
\vspace{-0.5cm}
\caption{The sensitivity of $R=\sigma_L/\sigma_T$ for
the $\rho$ production to the S-D-mixing \protect{\cite{Igor}},
for the compilation of the experimental data 
see \protect{\cite{ZEUSLT}}.}
\label{Figure4}
\vspace{-.3cm}
\end{figure}

The theoretical calculations \cite{Igor} seem to overpredict 
$R=\sigma_L/\sigma_T$  at large $Q^2$, see fig.~4,
for the compilation of the experimental data data see \cite{ZEUSLT}. On the 
one hand, the admissible
$S-D$ mixing brings the theory to a better agreement
with the data. on the other hand, as the recent data from ZEUS 
\cite{ZEUSLT}
do indicate, the experimental value of $R$ tends to rise with
the time. 
Here I would like to raise the issue of sensitivity of
$R$ to the short distance 
properties of vector mesons \cite{NNNCracow}. 

Consider $R_{el}= \sigma_{L}/\sigma_T$ for elastic CS 
$\gamma^*p\to \gamma^*p$, which is quadratic in the ratio of
CS amplitudes. By optical theorem one finds
\be
R_{el}= {\sigma_L(\gamma^*_L p\to \gamma^*_L p) \over 
\sigma_T(\gamma^*_T p\to \gamma^*_T p)}= \left|{A(\gamma^*_L p\to \gamma^*_L p) \over 
A(\gamma^*_T p\to \gamma^*_T p)}\right|^2= 
\left({\sigma_{L}\over \sigma_T}\right)^2_{DIS} \approx 4\cdot 10^{-2}
\label{eq:4.1}
\ee
Here I used the prediction \cite{NZHERA} for inclusive DIS
$R_{DIS} = \left.\sigma_{L}/ \sigma_T\right|_{DIS}\approx 0.2$, which
is consistent with the indirect experimental evaluations of
$R_{DIS}$ at HERA. 
Such a dramatic change from $R_{el}$ to $R$ of (\ref{eq:4.2})
suggests that predictions of $R$ for diffractive  VM
production are extremely sensitive to the poorly known
admixture of quasi-pointlike $q\bar{q}$ componets in VM. 

\section{Longitudinal gluons and polarization of a
direct $J/\Psi$ and $\Psi'$ at Tevatron}

There is a long standing mystery of the predominant 
longitudinal polarization of prompt $J\Psi$ and $\Psi'$
produced at large transverse momentum $p_{\perp}$ as
observed by the CDF collaboration in
inclusive $p\bar{p}$ interactions at Tevatron
\cite{CDF}, see fig.~5, in which the polarization
parameter'
$$
\alpha= {\sigma_T -2\sigma_L \over \sigma_T+\sigma_L}
$$
is shown (the observed $\Psi$' are arguably the direct ones,
the prompt $J/\Psi$'s include the  $J/\Psi$'s both
from the direct production and decays of higher charmonium 
states) . Specifically, the color-octet model
\cite{BraatenSigma} is able to parameterize the 
observed reaction cross section
(for a criticism of the standard formulation of 
the color-octet model see \cite{Regensburg}), 
but fails badly
in its predictions \cite{Leibovich,BraatenPolar} 
for the polarization parameter $\alpha$.

\begin{figure}[!htb]
   \centering
   \epsfig{file=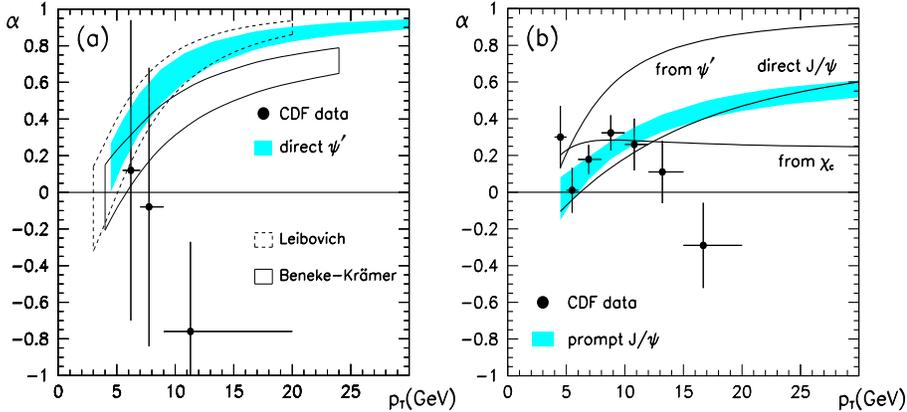,width=12cm}
\vspace{-0.3cm}
\caption{Predictions \protect{\cite{Leibovich,BraatenPolar}}
 from the color-octet 
model for the polarization parameter $\alpha$ vs. $p_{\perp}$ for
direct $\Psi$ and prompt $J/\Psi$ compared to CDF data
\protect{\cite{CDF}}.}
\label{Figure5}
\end{figure}

Arguably, production of charmonium states at mid-rapidity 
is controlled by gluon-gluon collisions. Now recall that 
in the standard collinear factorization the colliding 
gluons are regarded as on-mass shell, and transversely
polarized, ones. Which is the principal reason, why one
predicts the predominantly transverse polarization of
the produced $J/\Psi$ and $\Psi$'.

The QCD subprocesses for direct production of $C=-1$ vector 
states of charmonium are shown in fig.~6. In order to emphasize
an impact of their virtuality of the colliding gluons, I
indicate explicitly the origin of the gluon $g^{*}$. One
can readily show that, in close similarity to virtual
photons, the highly virtual gluons have both the familiar
transverse and so far ignored longitudinal polarization
\cite{JPsiPsiPrim}.

\begin{figure}[!htb]
   \centering
   \epsfig{file=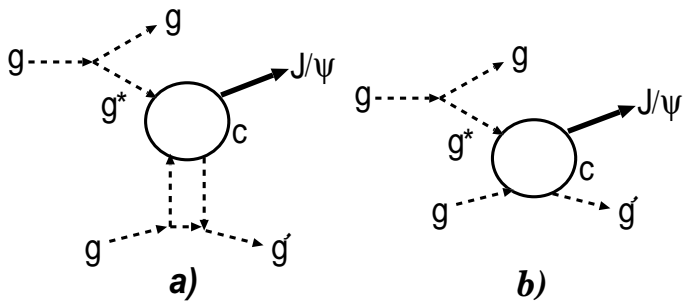,width=9cm}
\vspace{-0.3cm}
\caption{The diffractive QCD subprocesses for
the production of prompt vector states of
charmonium in hadronic reactions.}
\label{Figure6}
\vspace{-.3cm}
\end{figure}

In order to illustrate our principal point let me focus
on the "diffractive" mechanism of fig.~6a. 
It dominates at large invariant masses 
$\hat{w}$ of the $Vg'$ system, $\hat{w}^2 \gg M_{\Psi}^2$.
The virtuality of the gluon $g^*$ is
controlled by the transverse momentum $k_{\perp}$ of
the gluon $g^*$, so that $Q^2 \approx k_{\perp}^2$. 
The sub-process $g^* + g \to J/\Psi g'$
proceeds predominantly in the forward direction, which
implies that the transverse momentum of the $g^*$ is
transferred to the $J/\Psi$, so that
$p_{\perp} \approx k_{\perp}$.  The difference between color octet
two-gluon state in the $t$-channel of fig.~6a and
color-singlet two gluon state in the diffractive
pomeron exchange is completely irrelevant for spin
properties of the $J/\Psi$ production, and for the diffractive 
mechanism of fig.~6a we
unequivocally predict
$R =\sigma_L/\sigma_T \sim p_{\perp}^2/m_{\Psi}^2$,
i.e., $\alpha \to -1$ for very large $p_{\perp}$.
After some color algebra, one can readily relate the 
total cross section of the "diffractive" mechanism to
the cross section of photoproduction of $J/\Psi$ on
nucleons. We found that "diffractive" mechanism is
short of strength and could explain only $\sim 10$ per
cent of the observed yield of the direct $J/\Psi$. 

The diagrams of fig.~6b dominate for $\hat{w} \sim m_{\Psi}$.
Arguably, the above estimate for the $p_{\perp}$ dependence
of $R$ applied to this mechanism too. Crude estimates 
show that the contribution from this mechanism is commensurate 
to the "diffractive" production.

Besides the predominantly "forward" production when the
transverse momentum of the $g^*$ is transferred predominantly
to the direct $J/\Psi$, one must also consider the 
large angle reaction $g^* g \to J/\Psi + g$, which could affect
the polarization parameter $\alpha$. The full 
numerical analysis has not been completed yet, still 
we believe that the so far neglected longitudinal gluons
resolve a riddle of the longitudinal
polarization of direct $J\Psi$ and $\Psi'$.

\section{Conclusions }

QCD theory of diffractive production of vector mesons is
in a good shape and offers a solid basis for the quantitative
interpretation of the experimental data. So far neglected
longitudinal gluons are predicted to dominate production
of direct vector mesons at large transverse momentum
in hadronic collisions and resolve the long-standing
riddle of the dominant longitudinal polarization
of the $J/\Psi$ and $\Psi'$ discovered by CDF.

I'm grateful to my collaborators Igor Ivanov and
Wolfgang Sch\"afer for much insight and pleasure of
joint work on ideas reported here. 
Thanks are due to C. Ciofi degli Atti and M. Gianini
for invitation to this exciting
workshop.

\end{document}